\documentclass[aps,prb,eqsecnum,showpacs,draft,psfig]{revtex4}
\usepackage{amsmath,bm}

\usepackage{psfig}

\begin{document}

\title{Sound modes broadening for Fibonacci one dimensional
quasicrystals.}

\author{E.I.Kats} \affiliation {Laue-Langevin Institute, F-38042, Grenoble, France } \affiliation {L. D.
Landau Institute for Theoretical Physics, RAS, 117940 GSP-1, Moscow, Russia}
\author{A.R.Muratov} \affiliation {Laue-Langevin Institute, F-38042, Grenoble, France }
\affiliation {Institute for Oil and Gas Research, Moscow, Russia.}

\date{\today}

\begin{abstract}
We investigate vibrational excitation broadening in one dimensional 
Fibonacci model of quasicrystals (QCs). The chain is constructed
from particles with masses $1$ and $m$ following the Fibonacci inflation rule.
The eigenmode spectrum depends crucially on the mass ratio $m$. For $ m=3$
(what roughly mimics $AlPdMn$ icosahedral ($i$) QCs) there are three almost
dispersionless optic modes separated from the acoustic mode by three large gaps,
for $ m=1/3$ (what mimics $ZnMgY$ $i$-QCs) there is one
dispersionless optic mode and one acoustic mode. We calculate also the
eigenfunctions which near
the gaps are deviated strongly from simple (Bloch-like)  periodic functions.
All calculations performed self-consistently within the regular expansion over 
the three wave coupling constant. The problem can be treated as well
in the framework of the perturbation theory over
the
parameter $1-1/m$ which we formally consider as a small parameter.
The approach can be extended to three dimensional systems.
We find that in the intermediate range of mode coupling constants, three-wave broadening for the both
types of systems (1D Fibonacci and 3D $i$-QCs) depends universally
on frequency $\omega $ and scales as $c_1 \omega + c_2 \omega ^2$ (where $c_1$ and $c_2$ are constants).  
For smaller values of the coupling constant 
broadening turns out proportional to $\omega ^3$. 
Our general qualitative conclusion is that for a system with a non-simple 
elementary cell phonon spectrum broadening is 
always larger than for a system with a primitive cell (provided all other characteristics are the same).

\end{abstract}

 
\maketitle

\section{Introduction}
\label{int}
Extensive experimental and theoretical work on quasiperiodic systems
(i.e. materials which exhibit non-periodic
long-range order) has revealed interesting new properties which are not found
in either periodic or in disordered systems (see e.g., \cite{CS86} - \cite{SR03},
the references we will quote in more details in what follows).
One of the remarkable features of QCs, (e.g., revealed in neutron diffraction
and inelastic scattering experiments) is the apparent
conflict between the high structure quality of these materials and
their vibrational excitations which are rather reminiscent
of those for disordered materials \cite{CF03}, \cite{RF03}. 

Although as it was mentioned already above, there is a considerable literature 
discussing eigenmodes and
related
properties of QCs, and a number of sophisticated calculations have been published 
over the last 20 years, there is still a clear need for a simple (but yet non-trivial)
theoretical model with predictions which can be directly tested experimentally. 
For instance, results found by exact diagonalization of dynamical matrices at high symmetry points
of the small Brillouin zone for corresponding QCs approximants 
(see e.g. \cite{HK93})
lead to very rich density of vibrational states including 
many different modes. However there is still very limited
correspondence between these theoretical studies and only two or three
broad and almost dispersionless optical modes (besides acoustical phonons at smaller
frequencies) observed in experiments 
\cite{BB93}, \cite{BB95}, \cite{BB99}, \cite{DB99}, \cite{SK02}.
In part this frustrating situation is just due
to the lack of a simple and tractable model, and the main goal of
this paper is geared towards the building and testing of a simple model
for vibrational modes in $i$-QCs.
 
Our motivations for presenting this paper are twofold.
First is
based on a simple observation that 
a disallowed in conventional crystalline materials 5-th order rotational
symmetry of $i$-QCs determines the unique golden ratio $(\sqrt 5 - 1)/2$
of incommensurate length scales that completely defines the structure
of all $i$-QCs. As we will see in what follows that many robust and 
experimentally testable features of the
excitation spectra in $i$-QCs are sensitive to only this specific feature
of the structure. 
Second - we go one step further with respect to
the results known already in the literature
(see, e.g., \cite{KC96}, \cite{LO86}) providing in the present paper a systematic procedure for handling vibrational
mode broadening in QCs. The broadening is often considered as a nuisance, while as we will show,
it provides valued information on QC physical properties and eigenmode structure.
To our knowledge such a calculation of the eigenmode broadening in QCs was not carried out thus far.
The simplest model structure constructed by the same
golden ratio is the 1D Fibonacci chain.
Our aim is not to claim that the Fibonacci 1D model we propose necessarily holds
for real 3D $i$-QC materials but to analyze the model and to compare qualitatively 
its predictions with
experimental data.
Note, however, that all calculations can be performed self-consistently not only in 1D 
but as well can be generalized for 3D QCs.
Besides we study the model within the regular expansion over the parameter $\epsilon = 1 - 1/m$,
assuming that $\epsilon \ll 1$, where $m$ is some ''effective binary QC''(see its definition below in the section \ref{mod}) mass ratio.
Of course our oversimplified model is only a zero approximation which 
may not have exactly right numbers but
could lead
to more or less right shapes of the dispersion laws.                                                            
The reasonability of our assumptions, model and physics behind will be commented at various places 
in this paper.

Our paper is organized as follows.
In the next section \ref{mod} we describe our model and calculate
its vibrational spectrum. In section \ref{eig} we compute the
eigenfunctions, and in section \ref{three} we find the mode
broadening due to anharmonic three wave coupling.
Finally we review and discuss our results in conclusion section \ref{con}.
In two appendices to the paper we
collect some more specialized technical material required for the calculations 
phonon line broadening due to anharmonic third order processes 
for the 5-particle approximants to the infinite Fibonacci chain (appendix \ref{A})
and three wave phonon broadening in isotropic 3D system with $q$-space limited by the sphere
$|{\bf {q}}| = q_0$ (appendix \ref{B}).
Those readers who are not very interested
in mathematical derivations can skip
these appendices finding
all essential physical results in the main text of the paper.

\section{Fibonacci Model}
\label{mod}
As it is well known a regular periodic 1D lattice can be generated from
one basic unit cell by simple translation. 
For the ideal periodic system the solution of the equation of motion
is wavelike and the vibrational spectrum forms one or more vibrational bands.
The density of state is singular near these bands edges. In the opposite limit
of the totally disordered lattice, the wave functions exhibit localization
behavior, and one has only a discrete spectrum.
The quasiperiodic lattices we are interested in this paper
are intermediate in this sense between ideally periodic and totally disordered systems.
To generate 1D quasiperiodic
system one has to apply more general procedure.
We are constructing the chain from particles with masses $1$ and $m$
in the following way
(so-called the Fibonacci inflation
rule):
\begin{eqnarray}
\label{a1}
(1)\, ,\, (1,m)\, , \,(1,m,1)\, , \, (1,m,1,1,m)\, ,...\,,
\end{eqnarray}
and so forth such that the ratio [number of $m$]/[number of $1$] approaching
the golden mean value $\sigma \equiv (\sqrt 5 - 1)/2 \simeq 0.62$ (the same as for $i$-QCs)
in the limit of long
sequences.
A symmetrical inflation rule $ 1 \to m$ and $m \to m\, 1$ can be also used to generate a Fibonacci chain
where the roles of majority and minority basis are exchanged.
To determine the phonon spectrum of the Fibonacci chain we have to solve
the dynamic equations for every atom ($u_i$ is the displacement of the $i$-th atom):
\begin{eqnarray}
\label{a2}
m_i \frac {d^2 u_i}{dt^2}= 2u_i-u_{i-1}-u_{i+1} \, . 
\end{eqnarray}
Here and below we employ the units with elastic moduli equal to unity.
Of course it is impossible to solve (\ref{a2}) analytically for an arbitrary chain,
with more than 4 particles, 
therefore numerical methods should be used to investigate the system.

The computations can be performed in many ways.
For example one  can find the vibrational spectrum using the standard transfer
matrix formalism \cite{LO86}.
However, predominantly aiming to analyze mode broadening, we will use here another approach
seems to be more appropriate for our purposes.
First of all it is more convenient instead of the infinite Fibonacci chain to study the
finite chain of the 
length $N$ with zero boundary conditions for the displacements $u(N)=u(0)=0$,
and to have the spectrum for the infinite chain, the results
should be computed in the limit $N\to \infty $. For the conventional
1D crystal, i.e. 
$m=1$ we easily obtain the well-known spectrum
\begin{eqnarray}
\label{a3}
\omega (n)=2 sin(2 \pi n/N)\ .
\end{eqnarray}
In the long wavelength limit, $\omega \to 0$, the details of the quasiperiodic
structure do not play a significant role for vibrational spectra (unlike electronic spectra),
and therefore we get in this limit the same spectrum (\ref{a3}). Evidently it is not the case for finite $\omega $
and $m \neq 1$. Let us chose $m$ to be more specific and somehow in contact with real $i$-QCs.
Almost all known $i$-QCs are three component alloys, for
example, $Al_{68}^{27}Pd_{21}^{106}Mn_{11}^{55}$, $Zn_{60}^{65}Mg_{31}^{24}Y_{9}^{89}$,
where superscripts indicate the mass number of the element and subscripts
show the atomic concentration of the
element in the alloy. 
Unfortunately with the Fibonacci chain we can mimic simply only two component alloys. 
However, luckily the
mentioned
above $i$-QCs alloys roughly contain one more light component and two more heavy components, 
and for the both $i$-QCs we have the mass ratio between the average heavy 
and the light component about $1:3.3$ at the composition
$68:32$ for $AlPdMn$, and the mass ratio is $1:2.8$ at the composition $31:69$ for $ZnMgY$.
We conclude that the composition in the both cases is not very different 
from the golden ratio, therefore 
the Fibonacci chain approximation at least in this respect might serve quite reasonable.

To have a nontrivial solution for the particle displacements $u_i$, the
eigenmode frequency $\omega $ must satisfy the eigenvalue equation.
The latter one, e.g., in the transfer matrix technique, is the condition
that the determinant of the product of $N$ transfer matrices should be 
zero. It allows to find easily the phonon spectrum of the Fibonacci chain
(see e.g. \cite{KC96}, \cite{LO86}).
The spectrum has gaps which may be labeled by the so-called Bloch index
$\kappa _s = (1/2) s \sigma \, (\rm {mod}.\, 1)$
($\sigma $ is the golden mean ratio) and the size of the gaps decreases
roughly with increasing $s$.
The model exhibits characteristics of both
a regular periodic and a disordered
system. In the low-frequency region, the system
behaves as a regular periodic crystals (and the vibrational eigenfunctions
appear extended), in the high-frequency region, there is no unique behavior 
for the eigenfunctions, and the spectrum shows many gaps.
However the exact solution is, so to speak, too exact for our purposes, and contains
too many subtle details of the model
such as the hierarchical nature of gaps, and
all branches of excitations simultaneously, while experimentally observed spectra
measured at room temperature
are much more poor, and it is not clear whether it is possible at all
to observe or to test experimentally these theoretically predicted
features of the spectra and to find somehow their characteristics.
For $i$-QCs  $Al Pd Mn$ and $Zn Mg Y$,
for which detailed studies have been carried out
\cite{BB93}, \cite{BB95}, \cite{BB99}, \cite{DB99}, \cite{SK02},
vibrational excitation spectra can be
separated into two well defined regimes:
the acoustic regime for frequencies smaller than, say, $6-8 meV$ and,
for larger frequencies, a regime in which the dynamical response is
characterized by a broad band of dispersionless optic-like modes.
The
optic like spectrum generally consists of 3 or 4 broad 'bands' (a few $meV$ wide), and no any gap
opening is observed.
That is all and, therefore, aiming to understand underlying basic physics
and to model even qualitatively observed dependences, one should not
refine the model to include some additional mechanisms and details,
but just to the contrary, one has to coarse - grained the model,
to have a benchmark to compare theoretical predictions and experimental data.

One more comment is in order here. Of course real $i$-QCs are not one-dimensional
Fibonacci chain. However the one-dimensional and three-dimensional
problems share a common mathematical foundation based on the golden mean ratio $\sigma $,
the both systems can be obtained by a projection method from a higher
dimensional space (two dimensional square lattice for the Fibonacci model \cite{KN84}),
and it is not surprising that they have common quite robust and generic
universal properties. As it is the case in related electronic problems
\cite{RM97}, \cite{RB03} we expect the qualitative features
of the spectra will carry over to two- and three-dimensional cases,
and similar results hold for any irrational $\sigma $.
Moreover phonon line broadening, our main concern in this paper,
is even more robust phenomenon in QCs than the spectrum itself.
Indeed for QCs we have deal with infinitely many density modulation
harmonics, filling densely the reciprocal space. However one always
can separate a major series of density modulation harmonics,
and the corresponding wave vectors scale like $\sigma ^n$, thus depending
mainly on universal QC building blocks ratio. Due to this fact
though formally the number of relevant states which are in resonance
with any given one is infinite, the most of these harmonics
have very small amplitudes, and therefore
their resonances are irrelevant in our coarse - grained approach.
The same can be seen by comparing characteristic time scale
of the processes responsible for the line - broadening.
For the particular density modulation, the time of energy
($\hbar \delta \omega $) transfer is of the order of $1/\delta \omega $.
Thus when the scattering is dominated by a fast three wave anharmonic
coupling we are investigating in this paper, there is no time
for an effective influence of the other resonances due to small harmonics
of the QC density modulation.
Although in this study we are mainly concerned with 1D systems, to support
aforesaid qualitative arguments we presented in the appendix \ref{B} 
the generalization of our approach to three dimensional systems.

Let us recapitulate the results of our analysis of the 1D Fibonacci chain vibrational
modes.
The following conclusions are deducted from the numerical solution
of the eigenvalue equation, we present in the Fig. 1.
\begin{figure}
{\centerline{\psfig{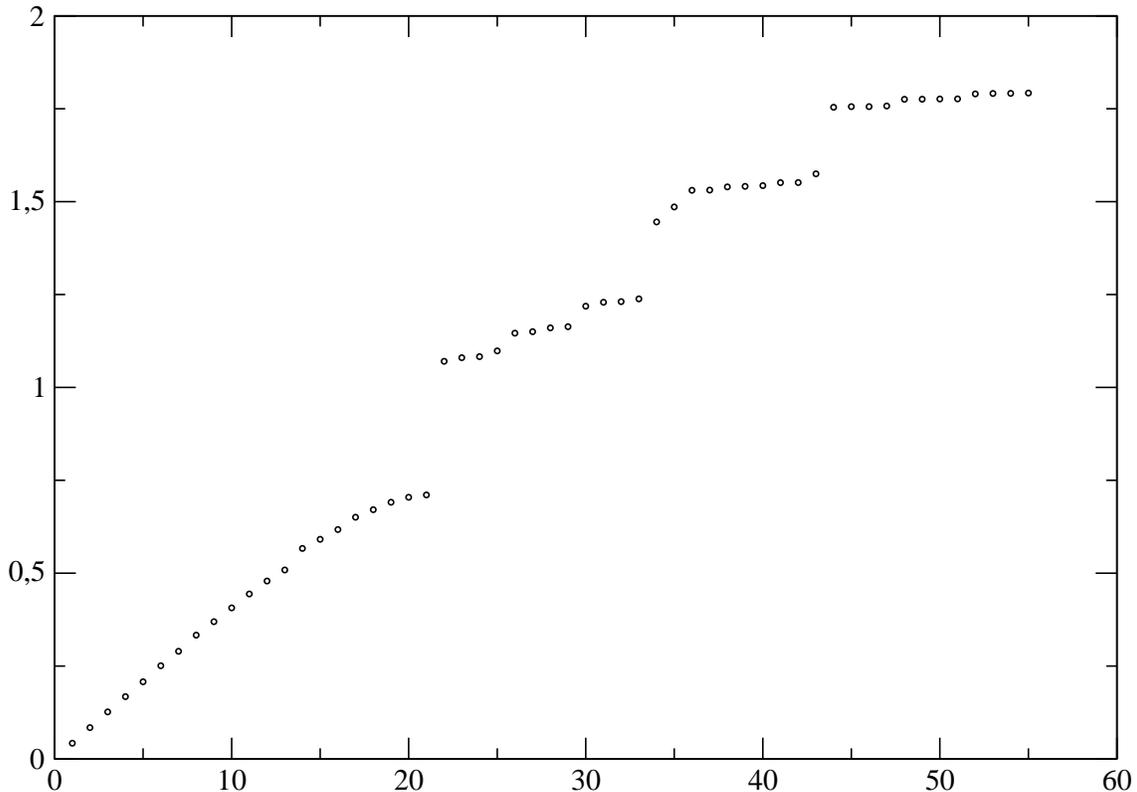} }} 
\caption{
Spectrum for the Fibonacci chain with 55 particles
in the elementary cell at zero displacement boundary conditions and 
for the mass ratio 1 : 3.} 
\end{figure}
The eigenmodes for 55 particles in the elementary cell shown in Fig.1, computed
with zero displacement boundary conditions and for the
mass ratio 1:3. Evidently one can distinguish one acoustic branch and optical branches.                                                
Qualitatively the same
features of the spectrum
occur also for 233 or 1000 particles.
Let us first consider the case when the mass ratio is
about 1:3 and the composition is about 38:62, which can mimic $i$-QC $AlPdMn$. 
By a simple coarse-grained (!) inspection of the results of such calculations we find 
that in this case we
have 3 optical modes with rather weak dispersion, one acoustic mode and one
''quasi-optical'' mode next to the acoustic one. The gap between the
acoustic branch and the first quasi-optical mode is relatively small, much smaller
than the other 3 gaps.
Analogously for the composition 62:38, the spectrum has only 2 branches, one acoustic 
and one optic mode, the latter one is almost dispersionless (Fig.2).
\begin{figure}
{\centerline{\psfig{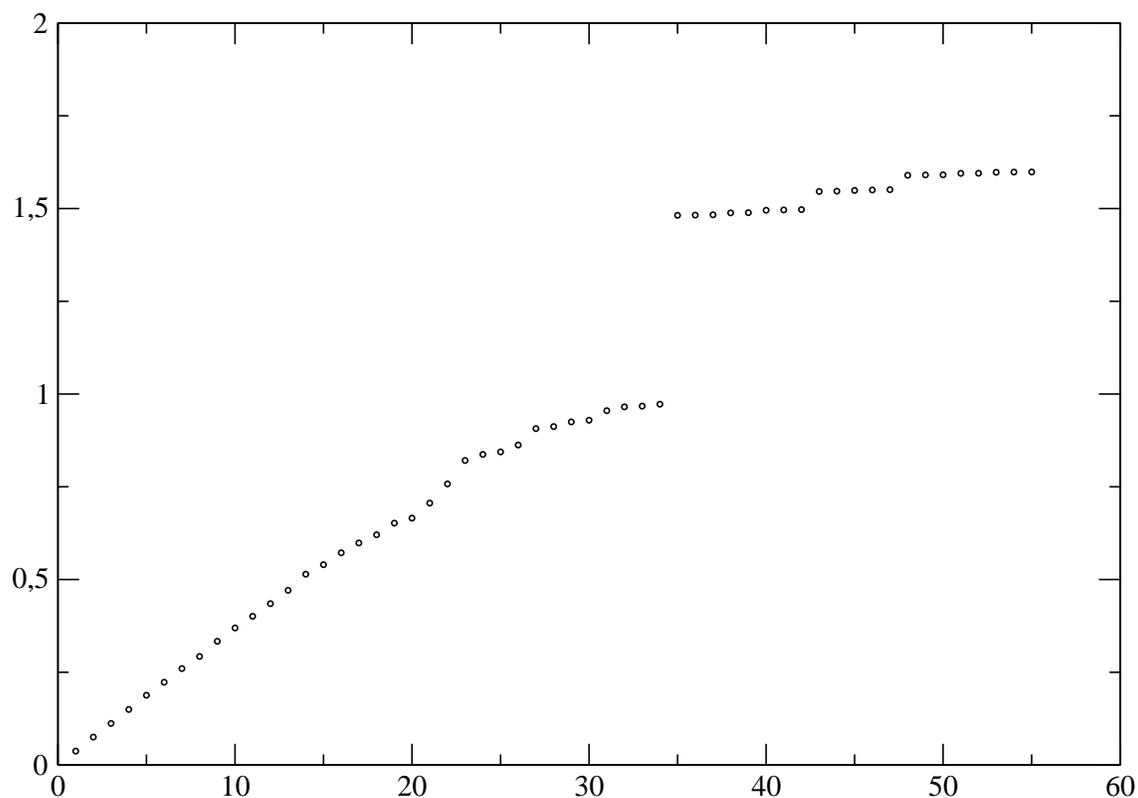} }} 
\caption{
Spectrum of the Fibonacci sequence for mass ratio $3 : 1$ (all other
conditions and parameters are the same as in Fig. 1).} 
\end{figure}
\par
The described above method (zero boundary displacements) 
widely used in the literature (see
e.g. \cite{KC96})
is adequate if one is looking for only the spectrum, however, the method is failed
for the consideration of anharmonic three wave contributions into mode broadening,
what is our calculations ultimate aim. It is evident in the limit $m=1$.
Indeed in the case the eigenfunctions are simply $sin$-functions and clearly
the mean product of any three $sin$
functions is always zero. 
For $m\neq 1$ this product is not zero any more, but it turns out very small.
Thus to analyze eigenfunctions and anharmonic contributions for the Fibonacci
chain one must use a different approach to describe vibrational modes.
For example we can take 5 (remind that 5 is the Fibonacci number)
particles in the elementary cell and impose non zero periodic boundary conditions, requiring
that phases of the displacements within the elementary cell, are confined in the interval
from $-\pi $ to $\pi $. In the standard solid state physics language it is the same 
as the introduction of quasi-momenta in the first Brillouin zone (evidently for 5 particles
in the elementary cell and periodic boundary conditions the introduced phase is
the quasi-momentum
up to
the constant factor). 
Luckily it occurs that the spectrum calculated in this simple
model in the harmonic approximation practically coincides with the spectrum obtained by
the previous method, although formally the both spectra are expressed in 
different variables: 
\begin{itemize}
\item 
in the former approach the spectrum consists from the discrete points; every mode
can be characterized by a certain number corresponding to a certain
mode numeration, (it could be, see, e.g., \cite{KC96}, the the eigenfunction nodes number),
and the spectrum is one per one correspondence between these numbers and
computed eigenfrequencies;
\item
in the latter approach the phase plays the role of a wave vector (which strictly speaking
is undefined in the previous approach since in the Fibonacci model there are no Bloch states).
\end{itemize}
Therefore now we have 5 modes (eigenfrequencies)
which can be represented as the functions 
of the phase or quasi-momentum.
The fact that the both spectra are fairly closed (see Fig.3)
is not an accidental coincidence. It is based on a described above natural
generalization of the notion of wave vectors for the quasiperiodic systems.
Even more we will see in the next section that in the limit of small frequencies
the eigenfunctions of the Fibonacci chain vibrational spectrum are practically
indistinguishable from the Bloch-like waves, thus one should expect that the
displacement phases (or the quasi-momenta)
are reminiscent of the mode numbers. 
\begin{figure}
{\centerline{\psfig{file=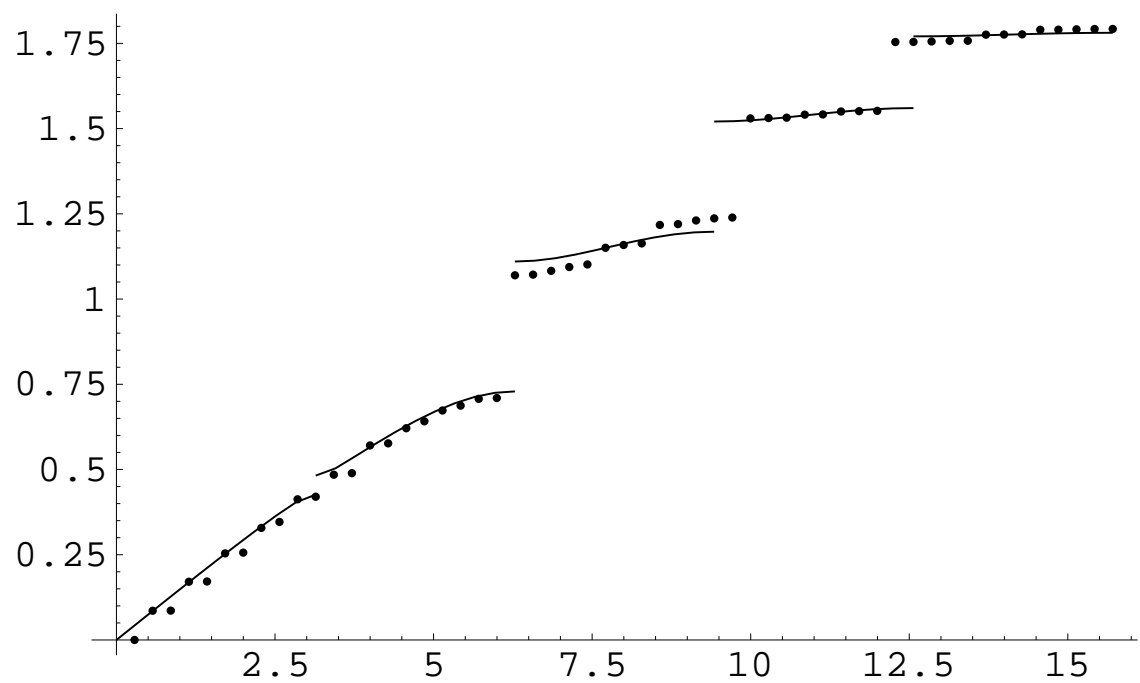,width=15cm} }} 
\caption{
Comparison of the eigen spectra computed with zero displacement
boundary conditions for the Fibonacci chain with 55 particles in the elementary cell
(points) and with periodic boundary conditions for 5 particles in the elementary cell
(solid line). The mass ratio is $1 : 3$ for the both cases (note,
that along the horizontal axis we put for the former case - mode number, and
for the latter case the phase or quasi-momentum).}
\end{figure}
A similar behavior is obtained for the next approximant in the series. As we proceed
by considering higher order approximant to the Fibonacci chain, new gaps progressively appear in the spectrum,
showing a hierarchical scaling structure. However, all corse-grained global features
of the spectrum remains the same as for very short approximants to infinite
quasiperiodic chains.

\section{Eigenfunctions}
\label{eig}

As it is well known if the system is periodic, Bloch's theorem may be applied, and
the solutions of (\ref{a2}) are wavelike, propagating ones. On the contrary,
if the lattice is disordered, the eigenfunctions are localized,
and the spectrum is a discrete set of levels. 
Common sense suggests that since the Fibonacci chain is intermediate
between periodic and disordered systems, it is expected to show the both characteristics.
Let us first consider the eigenfunctions for the discrete point-like spectrum
(i.e. the eigenfrequency as the functions of mode numbers). 
For the small eigenfrequencies the eigenfunctions are
similar to the Bloch-like periodic functions. For the particular case
of zero boundary conditions these eigenfunctions can be
written as $\sin (\pi n x/L)$. 
Near the gaps the eigenfunctions are represented as certain linear combinations
of the periodic functions, i.e., the wave packets.
The spectral width of any of the wave packet is proportional
to the value of the gap. The most broad eigenfunctions are at the boundaries of the
gaps, and the values of eigen numbers at the gaps are the corresponding Fibonacci numbers. 
The largest gaps occur for the mass ratio 1:3 at the positions after 2,3,4 elements 
(for the case when the unit cell length is 5). 
We illustrated some features of the eigenfunctions in the
Fig. 4 where we presented Fourier transforms of 5 typical eigenfunctions
for a chain with 55 particles in the elementary
cell. From this simple figure important conclusions are arrived at.
Namely as it was already noticed above for small eigen numbers 
the eigenfunctions only slightly differ from
periodic Bloch-like ones. However near the gaps the eigenfunctions
are strongly deviated from the Bloch solutions
and they describe localized or intermediate
(critical) states.
\begin{figure}
{\centerline{\psfig{file=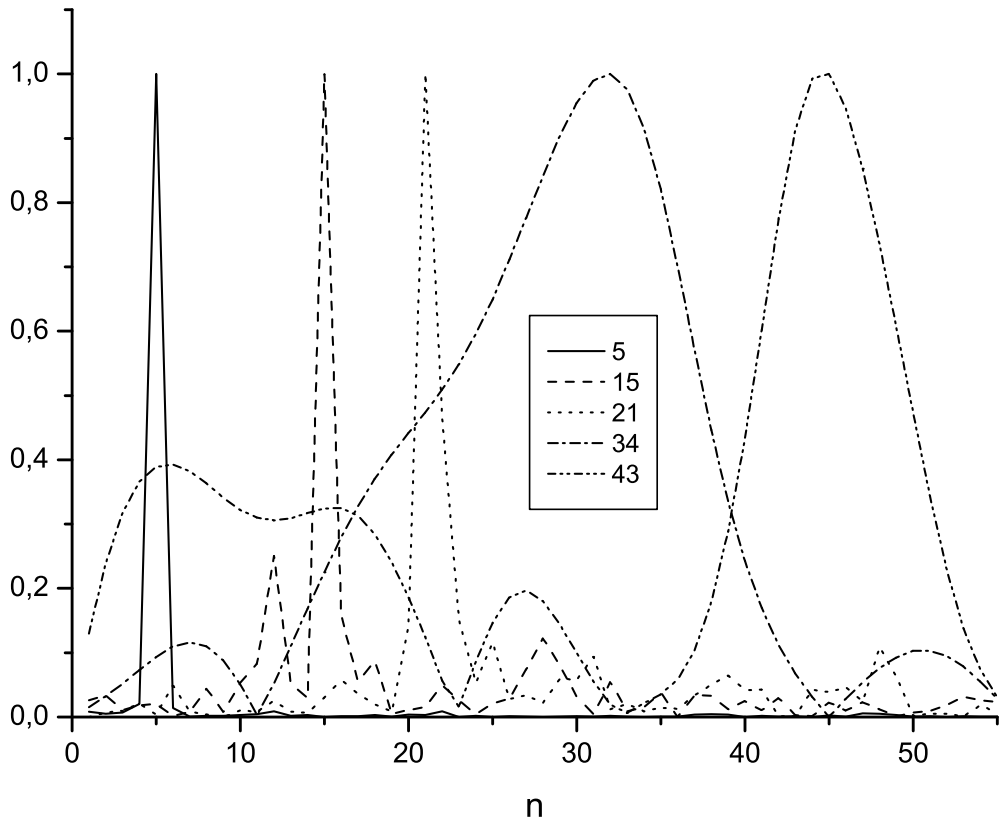,width=15cm} }} 
\caption{
Fourier transforms of eigenfunctions with mode numbers 5,15,21,34 and 43 
for the zero displacement boundary
conditions (the mass ratio is $1 : 3$).}
\end{figure}

To clarify further the situation few remarks are in order here:
\begin{itemize}
\item
Since the eigenfunctions are not Bloch-like periodic ones, 
in the thermodynamic limit, 
($N \to \infty $) there is no
the quasi-momentum conservation law;

\item
The eigenfunctions near the large gaps have quite broad widths  in the reciprocal ($k$) space, 
the spectral distributions of the neighboring eigenfunctions are strongly
overlapped. Having in mind neutron measurements it can lead to the 
situation when some phonon branches are disappeared at certain $k$ 
values and the next 
branch can appear at the lower quasi-momentum value; 

\item
We note that in all published reasonably accurate inelastic
single grain neutron scattering experiments, \cite{BB93}, \cite{BB95}, \cite{BB99}, 
\cite{DB99}, \cite{SK02}, the predicted gaps in the vibrational spectra,
have not been detected. One can attribute this failure to the mode broadening,
which could be larger than the corresponding gaps.
However, say the natural broadening, related to the fact
that the calculated above eigenfunctions have a finite width in $k$ space, 
is rather small
and can not close the big gaps between the optical modes. 
Therefore one has to include the other broadening mechanisms,
and the most relevant one is associated with three phonon interactions
\cite{LL81}. 
\end{itemize}

\section{Three wave broadening}
\label{three}
As it was shown in the previous
section, that already 5 particles ordered as $1,m,1,1,m$ give the phonon 
spectrum qualitatively and semi-quantitatively quite close 
to the spectrum of the large Fibonacci chain.
Evidently,  for the 5 particles in the elementary cell
system, one has 5 branches of the excitations:
one acoustic branch (1), one optical mode (2) with non-zero dispersion close to the
acoustical mode, and three optical almost dispersionless (3,4,5) modes.
For the mass ratio $m=3$ the simple analysis of the spectrum
lead to the following not-forbidden anharmonic three phonon
processes:
$$
5\leftrightarrow 4+1\, ; \, 5\leftrightarrow 3+2\, ; \, 4\leftrightarrow 3+1\, ;\, 3\leftrightarrow 2+2\, ; \, 2\leftrightarrow 1+1 \, .
$$
Actually only some parts of the corresponding branches can participate in these
three phonon interactions. Indeed, for example, the processes with two phonons
from the dispersionless optical modes and one phonon from the acoustic or
from the first optical mode are allowed only for the small $k$ for the latter mode.
All other processes between such phonons are forbidden.
Thus we conclude that soft phonons with small wave vectors do not participate in 
these three phonon 
processes and, therefore, in this approximation such phonons have
no broadening at all. This observation is conformed with known results of
inelastic neutron scattering measurements close to the strong Bragg
peaks, which show that there is a characteristic wave 
vector $q_{1c}$ of the order of $(0.3 - 0.5) \AA
^{-1}$, such
that only for $q > q_{1c}$ the sound modes exhibit broadening                                                       
\cite{BB93}, \cite{BB95}, \cite{BB99}, \cite{DB99}, \cite{SK02}.

\subsection{Basic model}

Just to illustrate the point and our calculation scheme, let us first consider the elementary cell with 2 atoms, 
$1$ and $m$. All allowed three wave processes depend strongly on the value of $m$.
For $m=3$ the only possible anharmonic  process is the decay of the optical phonon
with $k=0$ and $\omega =\omega_0$ into the two phonons with $\omega=\omega_0/2$.
For smaller $m$ one has
a certain finite $k$ region (not only one point)
within the optical mode where this kind of three wave processes
is not forbidden.
To study the three wave phenomena more rigorously one should start
with the vibrational Hamiltonian of the system including third order
anharmonic terms. The Hamiltonian can be written as \cite{LL81}
\begin{eqnarray}
\label{c1}
H=\sum_s \int dk \bigl( a_s^*(k,t)\omega _s(k)a_s(k,t)+\sum_{s_1,s_2}\int dk_1
V (a_s^*(k,t)a_{s_1}^*(k_1,t)a_{s_2}(k+k_1,t)+ c.c.) \bigr) 
\end{eqnarray}
Here $a_s^* \, ,\, a_s$ are the creation or annihilation operators of the
corresponding phonons, $V$ is the three wave coupling potential,
the integration is performed over the
first Brillouin zone, $c.c.$ means the complex conjugated contribution, and the summation 
is performed over all branches of the phonon spectrum (for the simplest model with two particles in
the elementary 
cell, these are one acoustic and one optical branch).

We calculate the simplest one-loop correction to the phonon spectrum.
Since the expansion in the Hamiltonian (\ref{c1}) is over $\nabla u$ the triple phonon interaction
vertex $V$ can be presented  
as $V=\lambda \sqrt{\omega \omega_1 \omega_2}$. Note that actually the interaction vertex $\lambda $ has also 
some smooth dependence on the wave vectors which can be found only numerically. 
Therefore, to illustrate the essential physics by a simple picture, in the paper we neglect this dependence.

According to the general principles of quantum statistical physics
\cite{LL81}, the probability of a phonon decay, (i.e. the vibrational
mode broadening, we are interested in) can be calculated self-consistently
from the Hamiltonian (\ref{c1}). To do it one has to introduce
certain bare decrements ($\gamma _{1 \, , \, 2}$ for the simplest
model with two particles into the elementary cell) into the phonon eigenstates.
As a result we get the following self-consistency Born integral equations
\begin{eqnarray}
\label{c2}
&&\gamma_1(k,\omega_1 )=g\omega_1(k)\int dk^{\prime}
\frac{(\omega_2(k+k^{\prime})-\omega_1(k^{\prime}))
(\gamma_1(k^{\prime})+\gamma_2(k+k^{\prime}))}
{(\omega_1(k)+\omega_1(k^{\prime})-\omega_2(k+k^{\prime}))^2+
(\gamma_1(k^{\prime})+\gamma_2(k+k^{\prime}))^2}
\ ,
\\
\nonumber
&&\gamma_2(k,\omega_2 )=g\omega_2(k)\int dk^{\prime}
\frac{(\omega_1(k^{\prime})+\omega_1(k-k^{\prime}))
(\gamma_1(k^{\prime})+\gamma_1(k-k^{\prime}))/2}
{(\omega_2(k)-\omega_1(k^{\prime})-\omega_1(k-k^{\prime}))^2+
(\gamma_1(k^{\prime})+\gamma_1(k-k^{\prime}))^2}
\ ,
\end{eqnarray}
where $g \propto T\lambda^2$.
Our main concern here is the case where the phonon broadening is
rather large, therefore the thermal phonon excitations giving dominate
contributions into the broadening should be thermally occupied, i.e.
$T \geq T_D$ ($T_D$ is Debye temperature), as it is the case in the experimental 
inelastic neutron scattering investigations \cite{BB95}, \cite{DB99}, \cite{SK02}
we have in mind in this paper. Thus in the condition $T \geq T_D$ we replace
the Bose energy level occupation factor
$(\exp (\hbar \omega /T) - 1)^{-1}$ by 
the Boltzmann distribution function 
$T/\hbar \omega $ in the high temperature limit. 
The solutions of these equations can be easily found numerically, and we presented
the results in Fig 5.
\begin{figure}
{\centerline{\psfig{file=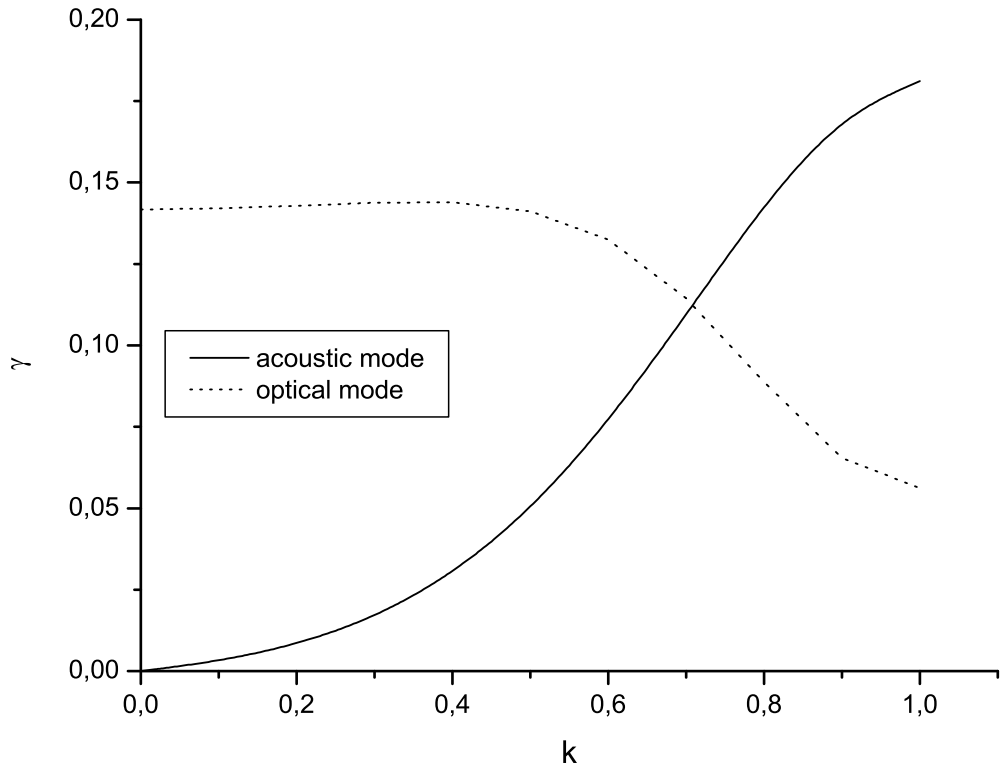,width=15cm} }} 
\caption{Phonon broadening 
calculated self-consistently from (\ref{c2}) (the mass ratio is $1 : 2.81$ and the
three wave interaction vertex $g=0.05$.}
\end{figure}
One might think that the equations (\ref{c2}) do not  contain any QC-specific feature
of the system, but it is only partially true. In fact 
the mode broadening (\ref{c2}) is determined by the key QC 
property, namely,
a finite band of almost dispersionless optic modes
interacting with the acoustic phonon.
On the same footing the aforesaid statement is applied 
in our approach, to any system with a non-simple unit cell.
Indeed the low-lying optical modes from approximant crystals might be
practically indistinguishable from quasi-local modes. 
The broadening depends
(at a given mass ratio $m$) on the mode coupling parameter $\lambda $,
i.e. it is non universal.
It is worth noting also quite peculiar behavior for the optical mode broadening
decreasing for $k > 0.5$. The physics behind can be rationalized as follows.
For a given mass ratio about $2.8$ the conservation laws prohibit the three wave interactions of the type
$1+1\leftrightarrow 2$ for the optical phonons with the wave vectors $k>0.5$. This broadening
reduction phenomenom
is even more pronounced for the smaller coupling constant (as it is demonstrated below in Fig. 7).

To illustrate this non universality we show in 
Fig. 6 the calculated three wave broadening 
for a relatively large anharmonic coupling ($g = 0.05$ in our
dimensionless units). 
\begin{figure}
{\centerline{\psfig{file=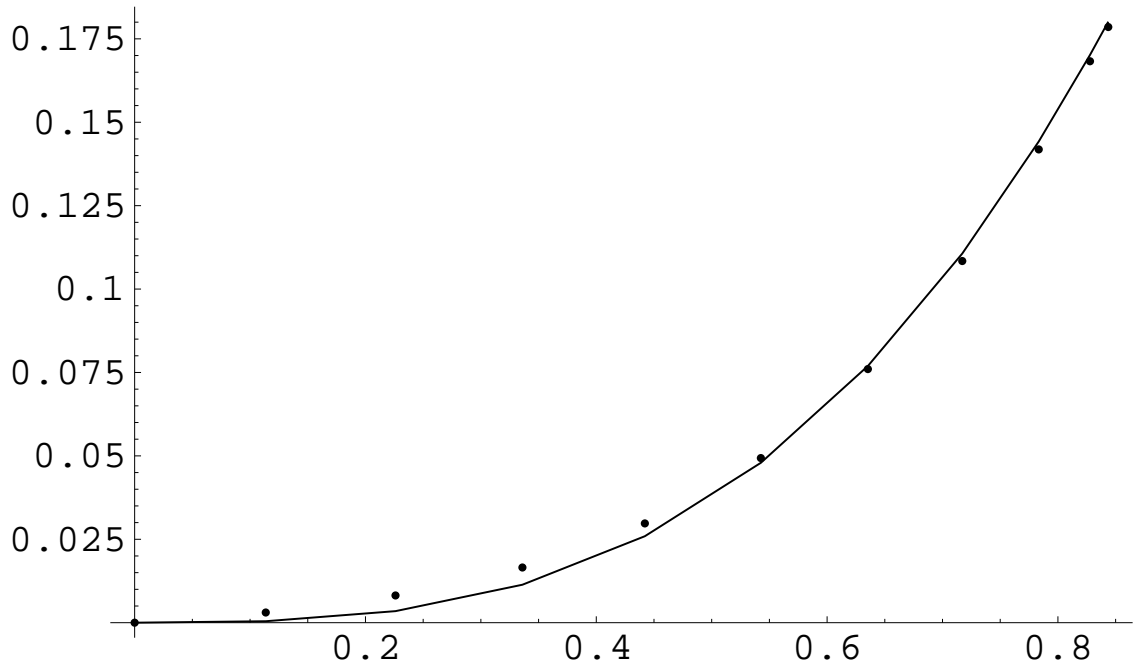,width=15cm} }} 
\caption{Phonon broadening as a function of $\omega $
(solid line is the function  $0.3 \omega ^3$, the parameters are the same as in Fig. 5).}
\end{figure}
The results in this case (strong coupling) can be fitted by $\omega ^3$ law. 
For a small coupling constant broadening law is quite
different, as we show in Fig. 7, where $g = 0.005$).
In the both cases the mass ratio was taken $1 : 2.81$, which mimics two kinds of $i$-QCs we described above. 
\begin{figure}
{\centerline{\psfig{file=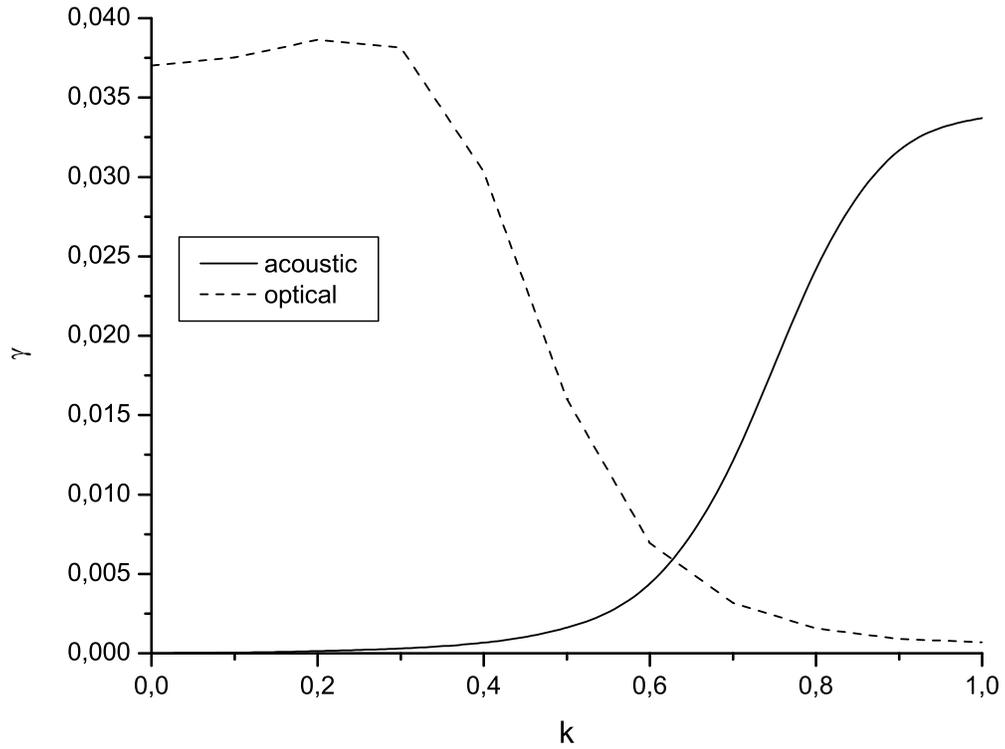,width=15cm} }} 
\caption{Phonon broadening 
calculated self-consistently from (\ref{c2})  (the mass ratio is $1 : 2.81$ and the
three wave interaction vertex $g=0.005$.}
\end{figure}
The figures 6 and 7 manifest clearly 
that $\omega ^3$ dependence does not hold for small anharmonic coupling, and
besides the figures show the difference between the regions 
of parameters corresponding to forbidden and allowed three wave processes.
It is worth noting that the three wave anharmonic coupling on the relevant in (\ref{c2}) length scales 
may be in fact much larger than anharmonic contributions 
conventionally measured by macroscopic (e.g. specific heat) methods
\cite{SF02}, \cite{SL02}).

We have shown already in section \ref{mod} that the global
structure of the vibration spectrum for quasiperiodic chains
can be obtained in practice by considering very short approximants
to infinite chains. For the Fibonacci chain, the reasonable size 
approximant imitating the infinite QC system contains
5 particle in the elementary cell, which is with our choice of masses
($m , 1 , m , m , 1$). Note in passing that the same statement
is true also for electronic spectra \cite{RM97}, \cite{RB03}.
Although the generalization  
of the presented above analysis 
and results for
two particles in the elementary cell to the unit cells with five
atoms is conceptually straightforward, 
it deserves some precaution, as it implies 
tedious and bulky
calculations, which could be done analytically only under certain rather
restrictive approximations. Appendix \ref{A} to the paper 
contains basic methodical details
and equations necessary for these calculations, 
and besides it gives a way to construct a regular method for calculating higher order
perturbative corrections. 
Here in Fig. 8 we present only the results
of our analysis for the 5-atom elementary cell.
\begin{figure}
{\centerline{\psfig{file=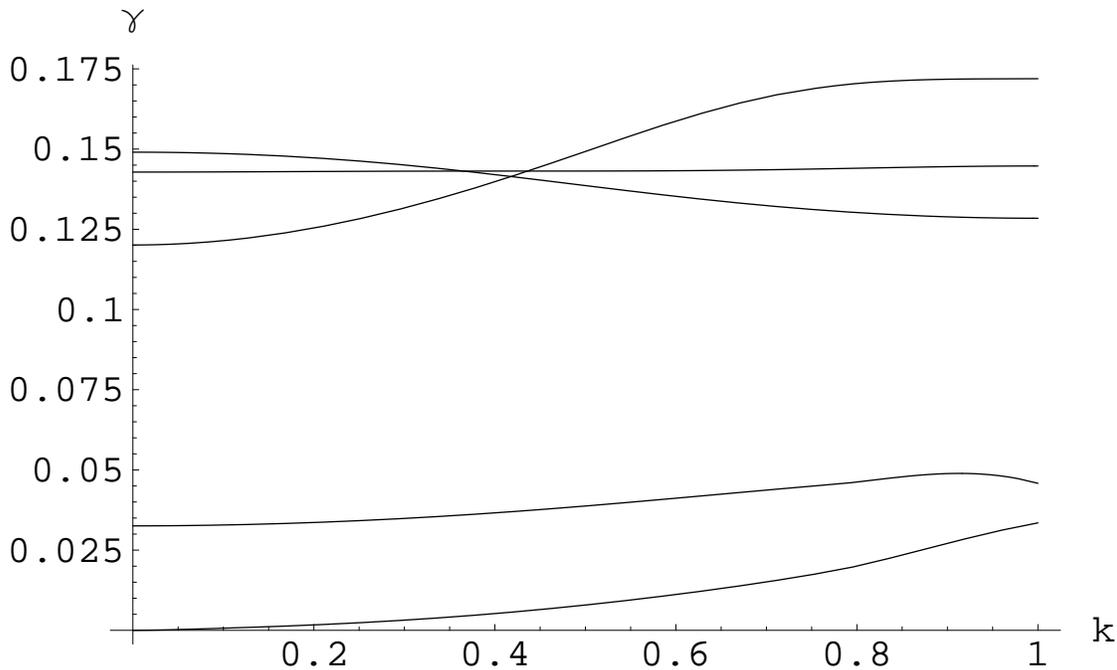,width=15cm} }} 
\caption{
Self-consistent solution of $(A1)$ for the mass ratio $1:2.81$ and
parameter $g=0.005$. Solid lines in this figure for $k=0$ are
branches with numbers $1, 2, 3, 4, 5$. The decrement for the lowest acoustic mode
tends to zero at $k \to 0$, the decrements for the higher energy branches $3,4,5$ are of
the same order of magnitude.}
\end{figure}
Only the main three-wave processes $1+1\leftrightarrow 2$, $2+2\leftrightarrow 3$, $1+3\leftrightarrow 4$, $1+4\leftrightarrow 5$, $2+3\leftrightarrow 5$
are taken into account at the computation, and besides, for the sake of simplicity
and having in mind published inelastic neutron scattering experimental data
\cite{BB93}, \cite{BB95}, \cite{BB99}, \cite{DB99}, \cite{SK02}, 
all in the range above Debye temperature,
we assume the classical
(Boltzmann)
statistics of the vibrational excitations.
 
Fig. 8 shows that the decrement for the lowest first acoustic branch tends to zero 
for small wave vectors (as it should be), and that the higher energy mode
broadening  $3,4,5$ are in the same range of magnitudes (of the order of $0.15$
in our dimensionless units).
Remembering that in the same units characteristic inter-mode
spacing (i.e. their eigenfrequency differences)
we found in section \ref{mod} is about $0.3$ we come to the following conclusions:
\begin{itemize}
\item 
due to the broadening 
in the range of resonances between acoustic and optic modes
the sound mode can no longer be described as a single excitation;
\item
large broadening of
the acoustic phonon modes 
is related to the three wave mechanism;

\item
the very existence of several quiet broad 
optical modes in QCs
can be understood as an illustration that in QCs
there are many ways in which the neighboring configurations
can be arranged, as a result a single level, which initially
was the same for all configurations becomes a band of the
modes;

\item
there are at least two main reasons why three phonon processes
lead to much more noticeable contributions to sound absorption in QCs 
(more precisely in the systems with non-simple unit cells) in comparison with conventional
crystals. First for the QC there is a dense set of umklapp vectors, and second
there are almost dispersionless optic modes possessing finite widths, and 
a large phase volume in the reciprocal space is available for the
acoustic phonons 
interacting with
the optical modes. 

\item
in own turn this noticeable broadening might be a reason
why no forbidden gaps
have been observed experimentally.                                            

\end{itemize}
Few more words on umklapp processes seem appropriate
here. As it is a textbook wisdom, the total momentum
of any set of interacting particles in a periodic crystal need only be conserved
to within a wave vector from the reciprocal lattice. The fundamental role of these so-called
umklapp processes is to open further scattering channels where the momentum of the
particles in the initial state is different from that of the final state.
In a QC its reciprocal space contains all the necessary wave vectors to match any required
frequency conversion processes. However this striking fact is almost irrelevant for the linear
phenomena since the corresponding susceptibilities anyway should remain constant in space.
Evidently it is not the case for non-linear susceptibilities and processes, like three wave
interaction, which as we illustrate in the frame work of our model can be enhanced noticeably 
in the quasiperiodic systems.   

\subsection{Refinement of the basic model} 
Presented in the previous subsection results can be extended in many ways.
First, for the sake of a skeptical reader we have to admit that strictly speaking our calculations
are not self consistent ones.
Indeed in our approach (\ref{c2}) (see also (\ref{apen1}) in the appendix \ref{A})
we have taken into account only three wave processes which are
not forbidden in the zero temperature limit.
By other words it means that there is no bare phonon mode broadening.
For finite temperatures and non-zero bare phonon decay,
the other processes could be also relevant, and the most important 
processes are $1+1\leftrightarrow 1$, because the spectrum of acoustic phonons is approximately
linear for small wave vectors (i.e., $\omega \sim c k$), which always allows three phonon interactions.
Therefore in the case of finite (and, in fact, not small) temperatures, we have to add the contributions
corresponding
to these $1+1\leftrightarrow 1$ processes into the right-hand side of the first Equation (\ref{c2}) (or for 5-particle
elementary cell into the r.h.s. of (\ref{apen1})).
We presented in the Fig. 9 the phonon broadening with such a contribution taken into
account (for the magnitude of the coupling vertex $g=0.005$).
\begin{figure}
{\centerline{\psfig{file=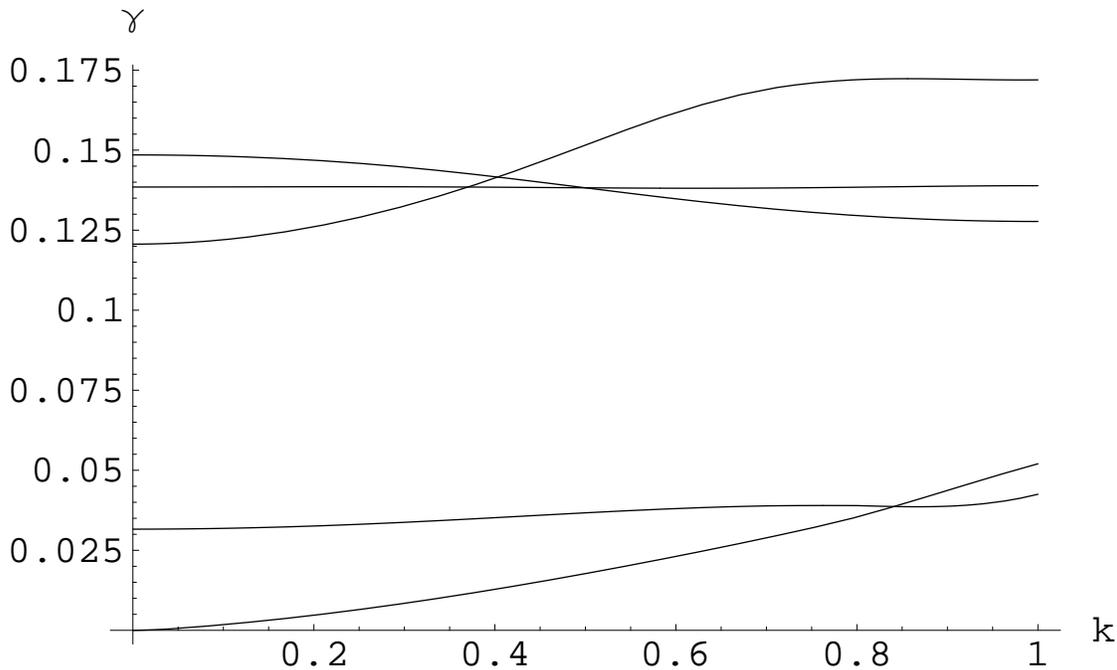,width=15cm} }} 
\caption{Vibrational mode broadenings for the $N=5$ approximant. The mass ratio is $1:2.81$ and
the coupling parameter is $g=0.005$.}
\end{figure}
A nice feature of this contribution is that the shape of the acoustic branch
decrement occurs almost independent of the magnitude of $g$ 
(for the interaction constant $g$ larger than the threshold which is about 0.005, see Fig. 10).
\begin{figure}
{\centerline{\psfig{file=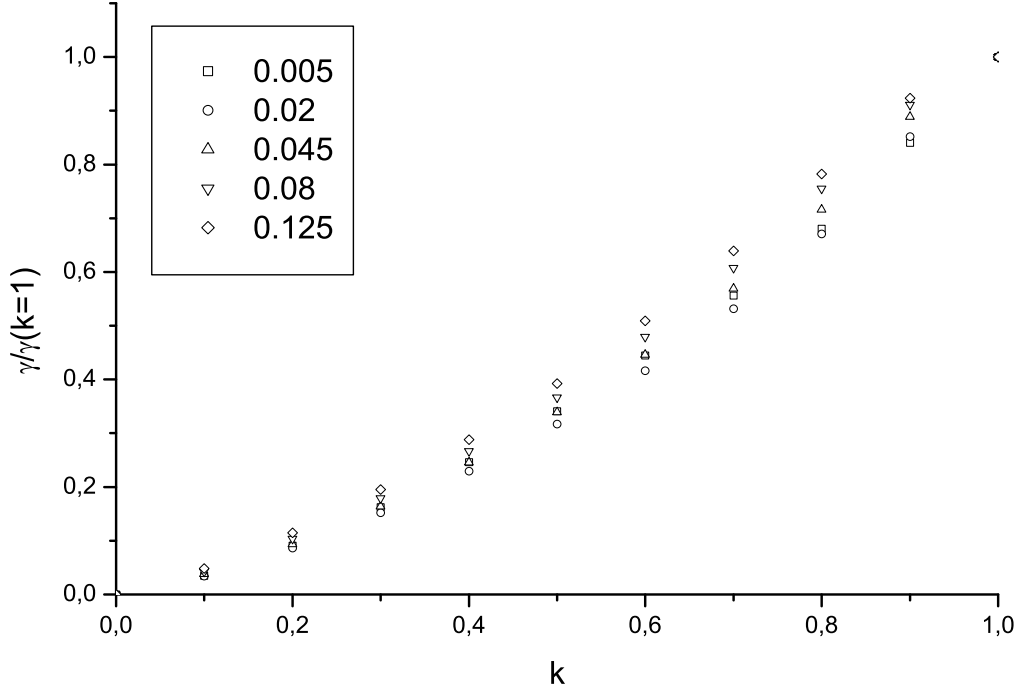,width=15cm} }} 
\caption{Acoustic mode broadening for the $N=5$ approximant. The mass ratio is $1:2.81$ and
the coupling parameter values are indicated in the figure.}
\end{figure}
One more note of caution is in order here.
In the previous subsection we replaced the Bose distribution function
$(\exp(\hbar\omega/T)-1)^{-1}$ 
by the classical occupation factor $T\hbar \omega $. Luckily the acoustic
branch broadening will not change its shape depicted in the Fig. 10 up to
the temperature larger than the maximum phonon energy in
the acoustic branch. Thus our approximation holds as well in a quite broad
range of temperatures.

This universal contribution
into the acoustic mode broadening can be also analyzed analytically,
and it illustrates several characteristic features of the phenomenom.
We believe that the phonon decrement is significantly smaller than its frequency.
To simplify consideration we assume also that the decrements of the optical modes do not depend on
$\omega $ or $k$, i.e. those aquire the constant values. Experimental data and our numerical investigations show
that optical phonon decrements indeed only very weakly depend on the wave vectors.
Replacing in the first equation in (\ref{apen1})  $\omega_1(k)\sim c |k|$,
($c$ as above is the sound speed), we are facing to solve
the integral equation
\begin{eqnarray}
&&\gamma_1(k )=g c k\int_{-1}^1 dq
\biggl( \frac{(\omega_2(k+q)-c|q|)(\gamma_1(q)+\gamma_2))}
{(c(|k|+|q|)-\omega_2(k+q))^2+(\gamma _1(q)+\gamma_2)^2}+  \nonumber
\\
&&\frac{(\omega_4(k+q)-\omega_3(q))(\gamma_3+\gamma_4)}
{(c|k|+\omega_3(q)-\omega_4(k+q))^2+(\gamma_3+\gamma_4)^2}+
\frac{(\omega_5(k+q)-\omega_4(q))(\gamma_4+\gamma_5)}
{(c|k|+\omega _4(q)-\omega_5(k+q))^2+(\gamma_4+\gamma_5)^2}+
\\
&&\frac{c(|k+q|-|q|)(\gamma_1(q)+\gamma_1(q+k))}
{c^2(|k|+|q|-|k+q|)^2+(\gamma_1(q)+\gamma_1(q+k))^2}+
\frac{c(|q|+|k-q|)(\gamma_1(q)+\gamma_1(k-q))/2}
{c^2(|k-q|+|q|-|k|)^2+(\gamma_1(q)+\gamma_1(k-q))^2}
\biggr) \nonumber
\end{eqnarray}
Since the decrement is assumed to be small in comparison with the frequency
it is possible to replace frequencies differences in the numerators in the following way
\begin{eqnarray}
&&\gamma_1(k )=g c k\int_{-1}^1 dq
\biggl( \frac {c|k|(\gamma_1(q)+\gamma_2))}
{(c(|k|+|q|)-\omega_2(k+q))^2+(\gamma _1(q)+\gamma_2)^2}+  
\label{t1}
\end{eqnarray}
\begin{eqnarray}
\nonumber
&&\frac{c|k|(\gamma_3+\gamma_4)}
{(c|k|+\omega_3(q)-\omega_4(k+q))^2+(\gamma_3+\gamma_4)^2}+
\frac{c|k|(\gamma_4+\gamma_5)}
{(c|k|+\omega _4(q)-\omega_5(k+q))^2+(\gamma_4+\gamma_5)^2}+
\end{eqnarray}
\begin{eqnarray}
\nonumber
&&\frac{c|k|(\gamma_1(q)+\gamma_1(q+k))}
{c^2(|k|+|q|-|k+q|)^2+(\gamma_1(q)+\gamma_1(q+k))^2}+
\frac{c|k|(\gamma_1(q)+\gamma_1(k-q))/2}
{c^2(|k-q|+|q|-|k|)^2+(\gamma_1(q)+\gamma_1(k-q))^2}
\biggr) 
\end{eqnarray}
Evidently the first three terms give the broadening proportional to $k^2$. The fourth term
for the positive $k$ values is determined mainly by the integration region $0<q<1$, where its denominator
is minimal. The last term can be transformed into the same form after the replacement
$q\to -q$. Eventually this equation can be rewritten as
\begin{eqnarray}
\nonumber
&&\gamma_1(k)=k^2 \biggl( c_2+c_1\int_0^1 dq
\frac{\gamma_1(q)+\gamma_1(q+k)}
{c^2(|k|+|q|-|k+q|)^2+(\gamma_1(q)+\gamma_1(q+k))^2}
\biggr) = 
\end{eqnarray}
\begin{eqnarray}
\label{t2}
&&=k^2 \biggl( c_2+c_1\int_0^1 dq
\frac 1{\gamma_1(q)+\gamma_1(q+k)} \biggr)
\simeq c_1 k + c_2 k^2
\ ,
\end{eqnarray}
where $c_1$ and $c_2$ are numerical coefficients. The last approximate equality in (\ref{t2})
is obtained after the first iteration in (\ref{t2}).
Thus  we end up with
the following universal analytical form for the acoustic mode decrement
$\gamma_1(k)=c_1k+c_2k^2$.  This simple expression fits very well computed numerically
function $\gamma_1 (k)$ (see Fig. 11).
\begin{figure}
{\centerline{\psfig{file=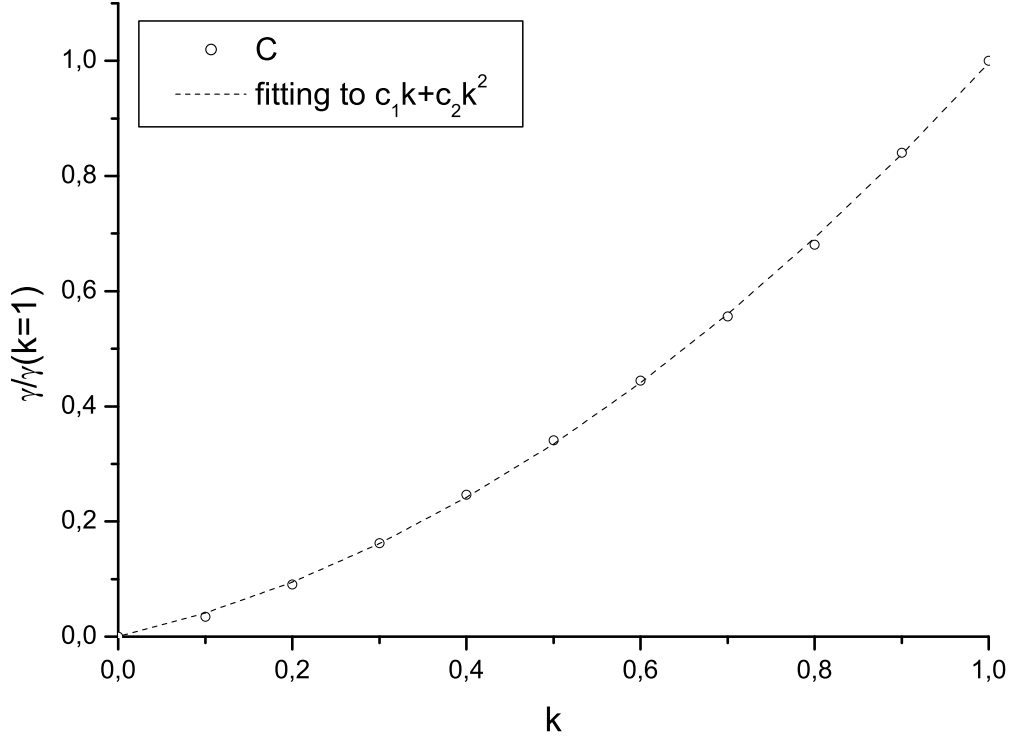,width=15cm} }} 
\caption{Numerically computed acoustic mode broadening for the $N=5$ approximant and its theoretical
fit by (\ref{t2}).}
\end{figure}
\newpage
In order to provide a more complete account of the vibrational broadening phenomena, similar analysis
has been performed to include self consistently the potentially relevant processes $n \leftrightarrow n+1$ ($n$ is the 
vibrational branch number, and as before $n=1$ corresponds to the acoustic mode). 
We closely follow the same procedure as above for the $1 \leftrightarrow 1 + 1$ broadening and thus skipping all details
presented only the results in Fig. 12.
\begin{figure}
{\centerline{\psfig{file=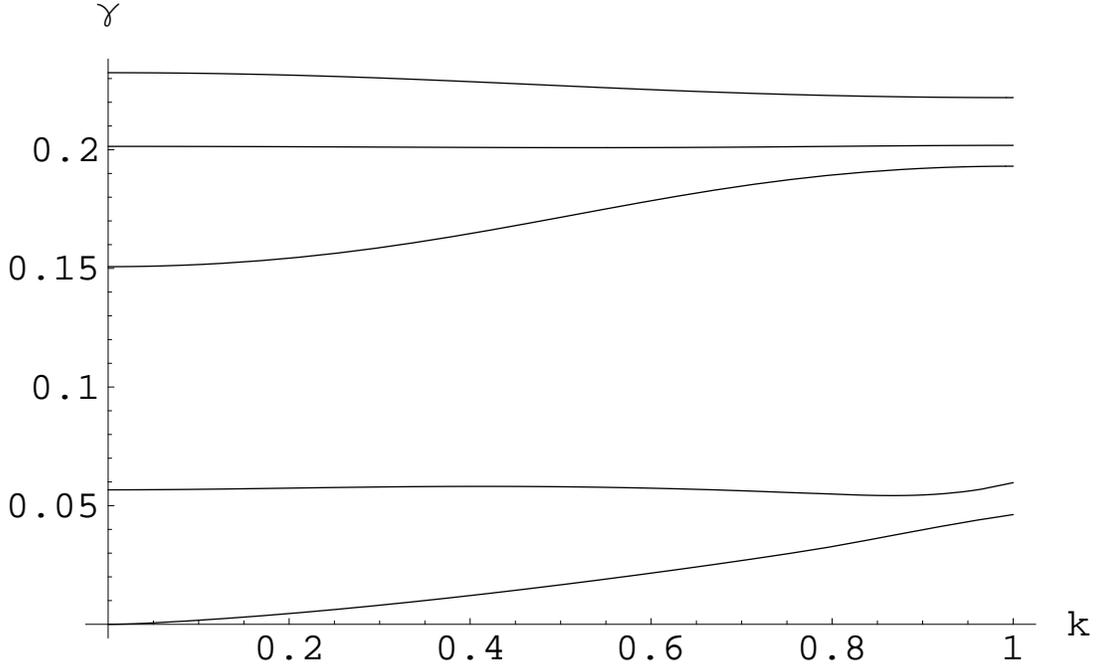,width=15cm} }} 
\caption{Vibrational mode broadenings with the processes $n \leftrightarrow n+ 1$ taken into
account ($N=5$ approximant and for the mass ratio $1:2.81$).}
\end{figure}
It turns out that these new processes have little effect on the acoustic branch decrement,
and the broadenings of all other branches only weakly depend on the 
wave vector and increase monotonically with the branch number.
A result that justifies the assumptions made above, and proves that our model includes all ingredients
necessary to capture the correct broadening effects in the Fibonacci chain.
\newpage

\subsection{Self consistent expansion over the parameter 
$\epsilon = 1 - 1/m$.}

Thus we investigated the Fibonacci lattice based on the golden ratio (and to confront
our results with known and well studied $i$-QCs the mass ratio was $m=3$) as an example
of the quasiperiodic 1D systems. There are many other non periodic structures
which are nonetheless fully deterministic and in this sense highly ordered.
Qualitatively the same behavior one can expect also in 1D incommensurate systems.
With this remark in mind it is interesting and instructive to 
extend the analysis of the previous section for a quasiperiodic system with the
mass ratio $m$ chosen to make use of the smallness                                                 
of the parameter $\epsilon = 1 - 1/m$.
If this assumption is granted we can apply a regular expansion over the small
$\epsilon $ what allows us to study even the infinite chain.

To move further on smoothly let us remind the characteristic features of the Fourier spectrum 
for periodically repeated
finite  Fibonacci block (remind that the length of the elementary cell
is a Fibonacci number also). 
According to the Nyquist theorem (see e.g., \cite{WE83}) for $N$
particles in the elementary cell one can determine only $N/2$ independent Fourier
harmonics. The absolute value of each of the Fourier amplitude is an even function of its wave number.
The zero wave number harmonic has the largest amplitude. The next large amplitude
harmonics have numbers $N_1$ and $N-N_1$, with the ratio $N_1/N$ close to
the golden ratio $N_1/N\sim 0.62$. The next by their amplitudes harmonics have the wave numbers 
$N_2\sim 0.62 N_1$ and so on. For example, for $N=21$ the wave numbers of the highest harmonics are $0, 8,
5,...$. Evidently the amplitudes of all harmonics with non-zero wave numbers are proportional to
the small parameter $\epsilon $, the corresponding proportionality coefficients are approximately constant and the
constant is smaller than one. The amplitude ratio of the different harmonics is independent 
of the small parameter $\epsilon $, and for the first two main harmonics (except the zero harmonic) this ratio is about 
$0.38$.

Armed with this knowledge we are in the position now to
consider the phonon spectrum in such a system. For $N$ particles in the elementary
cell we get the spectrum with one acoustic and $N-1$ optical branches.
The gaps between the branches are determined by the amplitudes of corresponding
Fourier harmonics. The amplitude of the harmonic, say, with the wave number  $K$ determines two spectral gaps with the numbers $K$
and $N-K$ respectively. Noting that the sum of these two gaps is approximately
equal to the amplitude of this harmonic, we arrive at the conclusion
that the vibrational spectrum of the Fibonacci chain
possesses the sequence of the gaps, and the value of the gap
depends on its wave number and the parameter $\epsilon $. In the limit $\epsilon \to 0$ all gaps also tend to zero.
The chain turns into the crystalline Bravais 
lattice with the only one particle in
elementary cell and the only one acoustic phonon branch.
For finite but small values of the parameter $\epsilon $
the difference
between the phonon spectra of the quasiperiodic system under consideration
and the corresponding Bravais lattice should be also small.
To be more specific let us focus on the simplest non Bravais system with three
particles in the elementary cell. The main feature of this non Bravais lattice 
is that the length of
Brillouin zone is three times smaller than for its Bravais counterpart, and besides
there are two additional non zero Fourier
harmonics. The acoustic branch of the non Bravais
lattice practically coincides with the corresponding part (one third) of the acoustic branch of
the Bravais lattice. However, the optical branches in the non Bravais lattice spectrum are 
significantly different. Let us denote by $q_0$ the size of the Brillouin zone for the Bravais lattice.
For
the optical modes we easily get that $\omega_2(k)\sim \omega_B(2q_0/3-k)$ and
$\omega_3(k)\sim \omega_B(k-2q_0/3)$, where $\omega_B$ is the only phonon frequency in
the Bravais lattice. Three wave interaction in the Bravais lattice is
determined by a certain vertex $\lambda_3$, which supposed to be not very different
from the similar triple 
interaction vertex 
between the acoustic phonons for the non Bravais lattice. However in the non Bravais case
we have to consider also all other interactions which include
the optical phonons. We denote the corresponding vertex $\lambda_4a(q_i)$, because
it describes in fact the four wave coupling, and $a(q_i)$ is the Fourier amplitude of the density modulation
at the wave vector $q_i$.
Actually the vertex $\lambda_3$ can also be reduced to $\lambda_4a(q_0)$,
and therefore the natural estimate for the vertices is $\lambda_4 a(q_i)/\lambda_3\sim a(q)/a(0)\sim \epsilon$.

The number of the wave vectors $q$ entering into the vibrational thermal broadening is equal to the number of the optical modes. 
Thus for our
simple case with three particles in the elementary cell we have deal with two vectors, and 
their Fourier amplitudes are equal and proportional to $\epsilon$.
Furthermore the effective triple vertex in the main over $\epsilon $ approximation and including all not forbidden
processes can be written as
$$\lambda_{eff}=\lambda_3\delta (k_1+k_2+k_3)+\sum_i \lambda_4 a(q_i)\delta (k_1+k_2+k_3+q_i)\ .$$
Finally the thermal vibrational mode broadening is determined by the square of this vertex. 
In a general case the different contributions into $\lambda _{eff}$ 
do not interfere.
Let us remind again that this conclusion is based on our assumption of $\epsilon \ll 1$,
what is equivalent to say that the amplitudes $a(q_i)$,
are small, and non small Fourier harmonics are included in the basic
structure (as we did above for the harmonic with the wave vector $q_0$).

Now we can compare the thermal broadening for the Bravais and non Bravais cases. The only three wave
process allowed in the Bravais lattice is evidently $1\leftrightarrow 1+1$. 
For the non-Bravais case
we have besides it also the three wave processes related to the interaction vertex
$\lambda_4 a(q_i)$. For our simplest case with three particles in the elementary cell
$a(q_0/3)=a(2q_0/3)$
and the both additional vertices coincide and proportional to $\epsilon$.
One can see that the vibrational broadening for the non Bravais case always exceeds the
broadening in its Bravais counterpart.
The corresponding difference of the both decrements $\gamma_3-\gamma_B$ is
proportional to the value $\sum_ia(q_i)^2$ (for the sake of simplicity we consider $\lambda_4$ to be a constant).
We present the results in Fig. 13 for the $\lambda_3=0.05$
and $\lambda_4=0.66 \lambda_3 \ (\epsilon \sim 2/3)$.
The qualitative inescapable message of this is that the phonon broadening is always larger for a non-simple elementary cell
(in comparison to the corresponding simple Bravais elementary cell).
\begin{figure}
{\centerline{\psfig{file=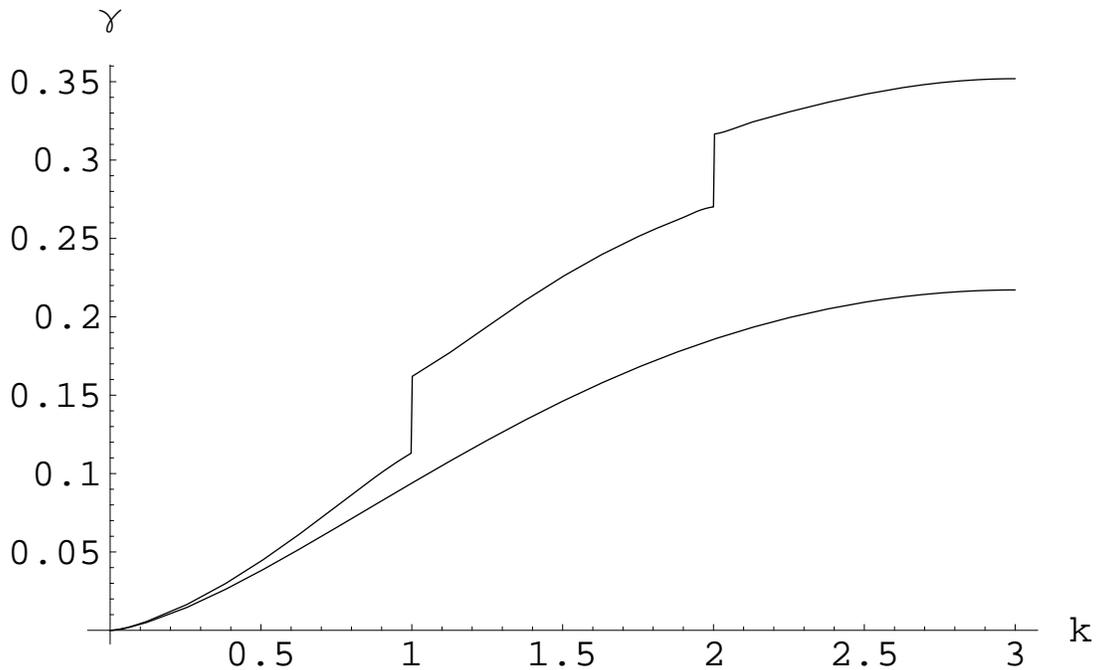,width=15cm} }} 
\caption{Comparison of the spectra for the Bravais system and its non Bravais counterpart
(the mass ratio is about $1:3$):
smooth line - the broadening in the 1D Bravais crystal;
curve with the jumps -
the broadening in the system with 3 particles in the elementary cell with the same
as for the smooth line interaction constant $g$.}
\end{figure}

\section{Conclusion}
\label{con}
In conclusion, in this paper we examine one important (and overlooked in previous
investigations) aspect of QC vibration spectra, namely - line broadening.
We concentrate on the broadening mechanisms which are intrinsic to
the structure of QCs, i.e. those that would be present at any dimension of QCs,
and even in a hypothetical defect-free materials.
To
understand physics, underlying the QC dynamics, we studied first the
simplest case - one-dimensional Fibonacci chain.
As it is well known facts
if the system is periodic, 
the Bloch theorem \cite{LL81} 
may be applied and,
therefore, the solution of the equation of motion
is wavelike, the phonon spectrum forms one or more bands,
and the density of states is singular near the band edges.    
On the other hand, if the lattice is totally disordered,
the eigenfunctions                                              
exhibits localization phenomena, and the spectrum is
a discrete set. The QC structure is intermediate between
between the ideal periodic and disordered, so it shows
characteristic of both systems.
For example in the low frequency limit $ \omega \to 0$,
the spectrum in our coarse-grained approximation appears as continuous
because the widths of gaps become smaller than the separation of
the eigenmodes in a chain of finite length.

Let us now add a few more ingredients en root to some tentative conclusions.
Our model chain is constructed
from particles with masses $1$ and $m$ following the Fibonacci inflation rule.
As the length of the pattern goes to infinity, the ratio between
the total number of elements of different components approaches constant
value.
The eigenmode spectrum depends crucially on the mass ratio $m$. For $ m=3$
(what roughly mimics $i$-QC $AlPdMn$) there are three almost
dispersionless optic modes separated from the acoustic mode by 
three large gaps. 
For $ m=1/3$ (what mimics $ZnMgY$ $i$-QCs) there is one
dispersionless optic mode and one acoustic mode. 
All calculations performed self-consistently within the regular expansion over 
the three wave coupling constant. The problem can be treated as well
in the framework of the perturbation theory over
the
parameter $1-1/m$ which we formally consider as a small parameter.
We find noticeable three wave anharmonic contributions into
the spectrum, i.e. to the mode broadening. The broadening depends
(at a given mass ratio $m$) on the mode coupling constant,
and  for relatively strong coupling, is proportional to $k^3$
($k$ is wave vector
transmitted to the mode). The need of an exponent lower than 4 for
$k$-dependent line broadening
(very often 
attributed to Rayleigh scattering in disordered materials
\cite{CF03}, \cite{RF03}) to fit the thermoconductivity data
in $Al Mn Pd$ $i$-QC was indicated also in \cite{KC96}.                                                                                                      
However, the line broadening dependence
on wave vectors turns out not universal one.
The acoustic mode decrement for the coupling parameter 
higher than $g = 0.005$ the exponent of the power law becomes smaller than 3, 
and for very small coupling this kind of power law fitting becomes inadequate.
We expect that found for 1D Fibonacci chain robust features of the vibrational
spectra will carry over to two and three dimensional cases. To illustrate this statement
we extended our approach also to three dimensional systems.
We find that in the intermediate range of mode coupling constants, three-wave broadening for the both
types of systems (1D Fibonacci and 3D $i$-QCs) depends universally
on frequency $\omega $ and scales as $c_1 \omega + c_2 \omega ^2$ (where $c_1$ and $c_2$ are constants).  
It is instructive to compare these predictions with the results known for say standard (i.e. possessing simple Bravais unit cells)
crystalline materials, where the inelastic anharmonic processes lead to a temperature dependent 
linewidth which scales as $k^2$, whereas elastic scattering of the Bloch like waves by the static
inhomogeneities leads to a temperature independent linewidth proportional to $k^4$ in 3D systems.
Our general qualitative conclusion is that for a system with a non-simple 
elementary cell phonon spectrum broadening is 
always larger than for a system with a primitive cell (provided all other characteristics are the same).
Although our model is a toy model in the sense of caricaturizing some physical
features (like, e.g., three wave broadening), when properly interpreted, it can yield quite reasonable values for a variety of measured
quantities. Our model establishes the universal properties of the vibrational spectra associated with 
generic features of any system with a non-simple
unit cell.

\subsection*{Acknowledgments} 
We thank 
M. de Boissieu and                               
R. Currat for the numerous discussions inspired this work. 
One of us (E.K.) acknowledges support from INTAS (under No. 01-0105) Grants.

\appendix
 
\section{}
\label{A}

The system of
equations for phonon line broadening in the 
5- particle approximant to the infinite Fibonacci chain,
reads   
\begin{eqnarray} 
\nonumber
&&
\gamma_1(k,\omega_1 )=g\omega_1(k)\int dk^{\prime } 
\biggl(
\frac{(\omega_2(k+k^{\prime})-\omega_1(k^{\prime})) 
(\gamma_1(k^{\prime})+\gamma_2(k+k^{\prime}))}
{(\omega_1(k)+\omega_1(k^{\prime})-\omega_2(k+k^{\prime}))^2+ 
(\gamma
_1(k^{\prime})+\gamma_2(k+k^{\prime}))^2}+ 
\end{eqnarray}
\begin{eqnarray} 
\nonumber
&&
\frac{(\omega_4(k+k^{\prime})-\omega_3(k^{\prime}))
(\gamma_3(k^{\prime})+\gamma_4(k+k^{\prime}))} {(\omega
_1(k)+\omega_3(k^{\prime})-\omega_4(k+k^{\prime}))^2+ (\gamma_3(k^{\prime})+\gamma_4(k+k^{\prime}))^2}+
\end{eqnarray}
\begin{eqnarray} 
\label{apen1}
&&
\frac{(\omega_5(k+k^{\prime})-\omega_4(k^{\prime})) (\gamma_4(k^{\prime})+\gamma_5(k+k^{\prime}))}
{(\omega _1(k)+\omega _4(k^{\prime})-\omega_5(k+k^{\prime}))^2 +
(\gamma_4(k^{\prime})+\gamma_5(k+k^{\prime}))^2} \biggr) 
\end{eqnarray}
\begin{eqnarray} 
\nonumber
&&
\gamma_2(k,\omega_2
)=g\omega_2(k)\int dk^{\prime}\biggl( \frac{\omega_1(k^{\prime})
(\gamma_1(k^{\prime})+\gamma_1(k-k^{\prime}))}
{(\omega_2(k)-\omega_1(k^{\prime})-\omega_1(k-k^{\prime}))^2+ 
(\gamma
_1(k^{\prime})+\gamma_1(k-k^{\prime}))^2}+ 
\end{eqnarray}
\begin{eqnarray} 
\nonumber
&&
\frac{(\omega_3(k+k^{\prime})-\omega_2(k^{\prime}))
(\gamma_2(k^{\prime})+\gamma_3(k+k^{\prime}))} {(\omega
_2(k)+\omega_2(k^{\prime})-\omega_3(k+k^{\prime}))^2+ (\gamma_2(k^{\prime})+\gamma_3(k+k^{\prime}))^2}+
\end{eqnarray}
\begin{eqnarray} 
\nonumber
&&
\frac{(\omega_5(k+k^{\prime})-\omega_3(k^{\prime})) (\gamma_3(k^{\prime})+\gamma_5(k+k^{\prime}))}
{(\omega _2(k)+\omega _3(k^{\prime})-\omega_5(k+k^{\prime}))^2+
(\gamma_3(k^{\prime})+\gamma_5(k+k^{\prime}))^2} \biggr) 
\end{eqnarray}
\begin{eqnarray} 
\nonumber
&&
\gamma_3(k,\omega_3
)=g\omega_3(k)\int dk^{\prime}\biggl( \frac{\omega_2(k^{\prime})  
(\gamma_2(k^{\prime})+\gamma_2(k-k^{\prime}))}
{(\omega_3(k)-\omega_2(k^{\prime})-\omega_2(k-k^{\prime}))^2 + 
(\gamma
_2(k^{\prime})+\gamma_2(k-k^{\prime}))^2}+ 
\end{eqnarray}
\begin{eqnarray} 
\nonumber
&&
\frac{(\omega_4(k+k^{\prime})-\omega_1(k^{\prime}))
(\gamma_1(k^{\prime})+\gamma_4(k+k^{\prime}))} {(\omega
_3(k)+\omega_1(k^{\prime})-\omega_4(k+k^{\prime}))^2+ (\gamma_1(k^{\prime})+
\gamma_4(k+k^{\prime}))^2}+
\end{eqnarray}
\begin{eqnarray} 
\nonumber
&&
\frac{(\omega_5(k+k^{\prime})-\omega_2(k^{\prime})) (\gamma_2(k^{\prime})+\gamma_5(k+k^{\prime}))}
{(\omega _3(k)+\omega _2(k^{\prime})-\omega_5(k+k^{\prime}))^2+
(\gamma_2(k^{\prime})+\gamma_5(k+k^{\prime}))^2} \biggr) 
\end{eqnarray}
\begin{eqnarray} 
\nonumber
&&
\gamma_4(k,\omega_4)=g\omega_4(k)\int dk^{\prime}\biggl(
\frac{(\omega_1(k^{\prime})+\omega_3(k-k^{\prime})) (\gamma_1(k^{\prime})+\gamma_3(k-k^{\prime}))}
{(\omega_4(k)-\omega_1(k^{\prime})-\omega_3(k-k^{\prime}))^2+ (\gamma
_1(k^{\prime})+\gamma_3(k-k^{\prime}))^2}+ 
\end{eqnarray}
\begin{eqnarray} 
\nonumber
&&
\frac{(\omega_5(k+k^{\prime})-\omega_1(k^{\prime}))
(\gamma_1(k^{\prime})+\gamma_5(k+k^{\prime}))} {(\omega
_4(k)+\omega_1(k^{\prime})-\omega_5(k+k^{\prime}))^2+ (\gamma_1(k^{\prime})+\gamma_5(k+k^{\prime}))^2}
\biggr) 
\end{eqnarray}
\begin{eqnarray} 
\nonumber
&&
\gamma_5(k,\omega_5)=g\omega_5(k)\int dk^{\prime}\biggl(
\frac{(\omega_1(k^{\prime})+\omega_4(k-k^{\prime})) (\gamma_1(k^{\prime})+\gamma_4(k-k^{\prime}))}
{(\omega_5(k)-\omega_1(k^{\prime})-\omega_4(k-k^{\prime}))^2+ 
(\gamma
_1(k^{\prime})+\gamma_4(k-k^{\prime}))^2}+ 
\end{eqnarray}
\begin{eqnarray} 
\nonumber
&&
\frac{(\omega_2(k^{\prime})+\omega_3(k-k^{\prime}))
(\gamma_2(k^{\prime})+\gamma_3(k-k^{\prime}))}
{(\omega_5(k)-\omega_2(k^{\prime})-\omega_3(k-k^{\prime}))^2+ (\gamma
_2(k^{\prime})+\gamma_3(k-k^{\prime}))^2} \biggr) 
\,  . 
\end{eqnarray} 

Here only the main three-wave processes $1+1\leftrightarrow 2$, $2+2\leftrightarrow 3$, $1+3\leftrightarrow 4$, $1+4\leftrightarrow 5$, $2+3\leftrightarrow 5$
are taken into account and besides we
assume classical statistics for the vibrational excitations.
Self-consistent solution to this set of equations shown in Fig. 8 for the mass ratio $1:2.81$ and
parameter $g=0.005$, qualitatively resembles experimentally observed spectra.
Note, however, that the line - broadening dependence
on wave vectors turns out not universal one.
The acoustic mode decrement for the nonlinear coupling parameter 
$g=0.005$ can be fitted fairly 
by the power law $\omega \propto k^3$, but for the higher values of $g$ the exponent
decreases, 
and for very small coupling power law fitting is not possible any more.

\section{}
\label{B}

Despite (at least partially) 
contradictory results of experimental investigations (s.f., e.g. \cite{BB99}, \cite{SK02}, and \cite{SF02})
a few conclusions 
on the following qualitative features of the phonon broadening in 3D systems
seem inescapable:
\begin{itemize}

\item 
Sound velocity is approximately isotropic;

\item The broadening for wave vectors $0.3-0.5 A^{-1}$ is also isotropic;

\item Phonon lines have almost Lorentzian shapes.

\end{itemize}
Note also that icosahedron and inverse dodecahedron (the main building blocks
for any $i$-QCs) are the most isotropic perfect polyhedrons.
The broadening due to three phonon interactions (we study in the main body of the paper)
is determined by the integral which
have singularity  along the line corresponding to the zero angle between the wave vectors.
All these features mean that the approximation which replace the first Brillouin zone by a sphere
should be quite reasonable one. Having this in mind we calculate the broadening
for the simplest case of the isotropic system with the reciprocal $q$-space limited by the sphere
$|{\bf {q}}| = q_0$ and with the only one phonon branch
\begin{eqnarray}
&&\gamma(q)={g \omega(q)^2}\int_0^{q_0} dk \, k^2 \int_{-1}^1 dt\,
\bigl( \frac {(\gamma({\bf k})+\gamma({\bf q}-{\bf k}))/2}
{(\omega({\bf q})-\omega({\bf k})-\omega({\bf q}-{\bf k}))^2+
(\gamma({\bf k})+\gamma({\bf q}-{\bf k}))^2} \nonumber \\
&&+\frac {\gamma({\bf k})+\gamma({\bf q}+{\bf k})}
{(\omega({\bf q})+\omega({\bf k})-\omega({\bf q}+{\bf k}))^2+
(\gamma({\bf k})+\gamma({\bf q}+{\bf k}))^2}
\bigr) \ .
\end{eqnarray}
Because everything is isotropic for the case, one has $\gamma({\bf k})=\gamma(k)$ and
$\gamma({\bf q}\pm {\bf k})=\gamma(\sqrt{q^2+k^2\mp 2qkt})$.
The integration above is
performed over the region $|{\bf k}|<q_0$. Thus we choose $q<q_0$,  and if
$|{\bf q}\pm {\bf k}|$ occurs to be larger than $q_0$, it must be replaced with
$2q_0-|{\bf q}\pm {\bf k}|$.
First term in r.h.s. corresponds to the decay of phonon with the wave vector ${\bf q}$,
the second term describes the fusion of this phonon with the phonon with the wave vector ${\bf k}$,
and to be specific the phonon spectrum is taken as $\omega(q)=\omega_0\sin (\frac {\pi q}{2q_0})$.
Of course for a more realistic situation it is necessary to consider at least
two particles in the elementary cell. The generalization is straightforward. In this case
the phonon spectrum consists from 3 acoustical and 3 optical branches, and due to the isotropy of
the system the transverse branches should be degenerate.
The solution of the corresponding equations can be performed as above and the results are presented in Fig. 14
(the eigenfrequencies) 
\begin{figure}
{\centerline{\psfig{file=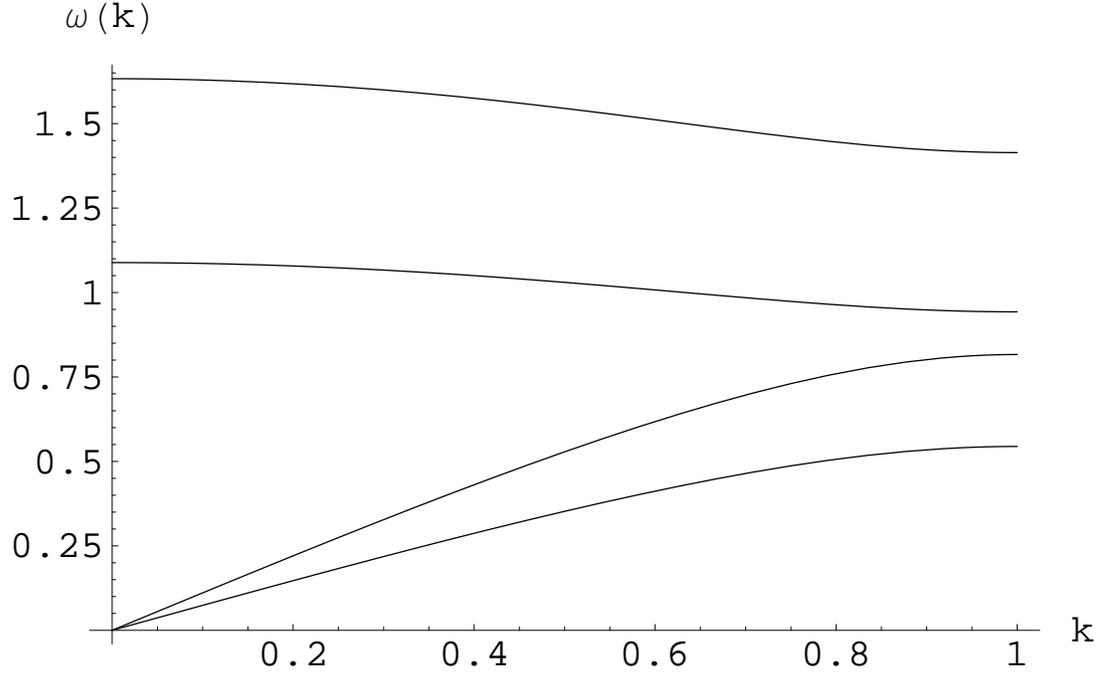,width=15cm} }} 
\caption{Phonon spectrum for the 3D isotropic model of $i$-QCs with three acoustic branches
and three optical branches (transverse modes are degenerate).}
\end{figure}
and in Fig. 15 (the vibrational mode broadenings). 
\begin{figure}
{\centerline{\psfig{file=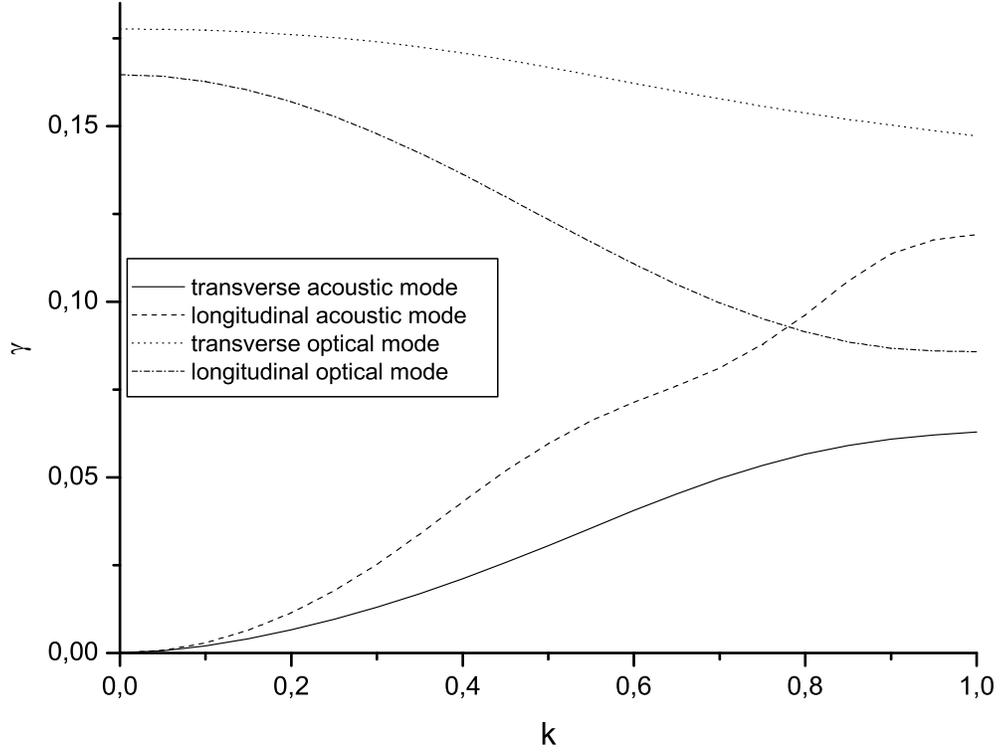,width=15cm} }} 
\caption{The broadening of the phonons for the same model as in the Fig. 14,
and for $\gamma/\omega\sim 0.16$.}
\end{figure}
Maximal frequencies for longitudinal acoustic and optical modes are $0.81$ and $1.4$
correspondingly, i.e. the ratio $\gamma/\omega(k=1)\sim 1/7$, i.e., quite close to 
the neutron scattering experimental data \cite{BB95}, \cite{DB99}, \cite{SK02}.
The robust qualitative features of the broadening 
for the both longitudinal and transverse acoustic branches occur very similar ones,
and can be fitted to $c_1k+c_2k^2$. This answer
is universal for the model, when the broadening is determined by only one
three wave interaction constant $g$.


\begin{references}
\bibitem{CS86} J.W.Cahn, D.Shechtman, D.Gratias, J.Mat. Res., {\bf 1}, 13 (1986). 
\bibitem{NI89}
K.Niizeki, J.Phys. A: Math. Gen., {\bf 22}, 4295 (1989). 
\bibitem{NA90} K.Niizeki, T.Akamatsu, J.Phys.:
Cond. Matter., {\bf 2}, 2759 (1990). 
\bibitem{JA92} C.Janot, Quasicrystals: a Primer, Oxford, Oxford Science (1992).
\bibitem{BB93} M. de Boissieu,
M.Boudard, R.Bellissent, M.Quilichini, B.Hennion, R.Currat, A.I.Goldman, C.Janot, J.Phys.: Condens.
Matter, {\bf 5}, 4945 (1993). 
\bibitem{HK93} J.Hafner, M.Krajci, J.Phys.: Condens. Matter, {\bf 5}, 2489 (1993). 
\bibitem{BB95} M.Boudard, M.de Boissieu, S.Kycia, A.I.Goldman, B.Hennion, R.Bellissent,
M.Quilichini, R.Currat, C.Janot, J.Phys.: Condens. Matter, {\bf 7}, 7299 (1995).
\bibitem{JA96} C.Janot, Phys. Rev. B, {\bf 53}, 181 (1996). 
\bibitem{KC96} P.A.Kalugin, M.A.Chernikov, A.Bianchi, H.R.Ott, Phys. Rev. B., {\bf 53}, 1445 (1996).
\bibitem{QJ97} M.Quilichini, T.Janssen, Rev. Mod. Phys., {\bf 69}, 277 (1997).
\bibitem{CO98}
M.A.Chernikov, H.R.Ott, A.Bianchi, A.Migliori, T.W.Darling, Phys. Rev. Lett., {\bf 80}, 321 (1998).
\bibitem{BB99}
R.Bellissent, M. de Boissieu, G.Coddens in Physical properties of quasicrystals, Z.M. Stadnik, ed.,
Springer (1999). 
\bibitem{DB99} F.Dugain, M.de Boissieu, K.Shibata, R.Currat, T.J.Sato, A.R.Kortan,
J.B.Suck, K.Hradil, F.Frey, A.P.Tsai, Eur. Phys. J. B, {\bf 7}, 513 (1999). 
\bibitem{JA00} T.Janssen, Ferroelectrics, {\bf 236}, 157 (2000). 
\bibitem{GS00} K.Gianno, A.V.
Sologubenko, M.A.Chernikov, H.R.Ott, Phys. Rev. B, {\bf 62}, 292 (2000). 
\bibitem{SK02} K.Shibata, R.Currat, M. de Boissieu, T.J.Sato, H.Takakura, A.P.Tsai, J.Phys.: Condens.
Matter, {\bf 14}, 1847 (2002). 
\bibitem{SF02} C.A.Swenson, I.R.Fisher, N.E.Anderson, Jr., P.C.Canfield, A.Migliori, Phys. Rev. B, {\bf
65}, 184206 (2002). 
\bibitem{SL02} C.A.Swenson, T.A.Lograsso, A.R.Ross, N.E.Anderson, Jr., Phys. Rev. B, {\bf 66}, 184206
(2002). 
\bibitem{KL03} M.Kleman, Eur. Phys. J., B, {\bf 31}, 315 (2003). 
\bibitem{SR03} V.V.Savkin, A.N.Rubtsov, T.Janssen, {\bf 31}, 525 (2003).
\bibitem{CF03} E.Courtens, M.Foret, B.Hehlen, B.Ruffl\'e, R.Vacher, J.Phys.: Condens. Mattter,
{\bf 15}, 1281 (2003). 
\bibitem{RF03} B.Ruffl\'e, M.Foret, E.Courtens, R.Vacher, G.Monaco, Phys. Rev.
Lett., {\bf 90}, 095502 (2003). 
\bibitem{LO86}
J.P.Lu, T.Odugaki, J.L.Birman, Phys. Rev. B, {\bf 33}, 4809 (1986).
\bibitem{LL81} L.D.Landau, E.M.Lifshits, Physical Kinetics (Course of Theoretical
Physics, volume 10), Pergamon Press, New York (1981). 
\bibitem{RM97} S.Roche, D.Mayou, Phys. Rev. Lett., {\bf 79}, 2518
(1997).  
\bibitem{RB03} S.Roche, D.Bicout, E.Macia, E.Kats, Phys. Rev. Lett., {\bf 91}, 228101 (2003).
\bibitem{KN84} P.Kramer, R.Neri, Acta Crys., A, {\bf 40}, 580 (1984).
\bibitem{WE83} H.J.Weaver, Applications of discrete and continuous Fourier
analysis, J.Wiley and Sons, New York (1983).

\end{references}
\end{document}